\documentclass[12pt,french,epsfig,graphics,twoside]{article}

\usepackage{graphicx}
\usepackage{amssymb}
\usepackage{textcomp}

\usepackage{cite}

\usepackage{bm}

\usepackage{mathabx}

\usepackage{dingbat}

\usepackage{pifont}

\usepackage{fancyhdr}
\usepackage[francais]{babel}
\usepackage[latin1]{inputenc}
\usepackage{tipa}
\usepackage{bbm}

\usepackage{graphicx}
\usepackage{amssymb}
\usepackage{textcomp}
\usepackage{stmaryrd}

\begin{document}

\newcommand{\nin}{\noindent}
\newcommand{\vv}{\vskip 0.25 cm}
\newcommand{\vvv}{\vskip 0.3 cm}
\newcommand{\vvvv}{\vskip 0.5 cm}

\newcommand{\beq}{\begin{equation}}
\newcommand{\enq}{\end{equation}}
\newcommand{\und}{\underline}
\newcommand{\biz}{\begin{itemize}}
\newcommand{\eiz}{\end{itemize}}
\newcommand{\di}{\displaystyle}
\newcommand{\bc}{\begin{center}}
\newcommand{\ec}{\end{center}}
\newcommand{\uta}{\begin{array}[t]{c}
{A'} \\
\stackrel{\sim}{\stackrel{~~}{~~}}
\end{array}}
\newcommand{\utx}{\begin{array}[t]{c}
{\xi} \\
\stackrel{\sim}{\stackrel{~~}{~~}}
\end{array}}
\newcommand{\pa}{\partial}
\newcommand{\g}{\gamma}
\newcommand{\ep}{\epsilon}
\newcommand{\cp}{\stackrel{\circ}{p}}
\newcommand{\ck}{\stackrel{\circ}{k}}

\newcommand{\uz}{\stackrel{\circ}{U}}
\newcommand{\az}{\stackrel{\circ}{A}}
\newcommand{\bz}{\stackrel{\circ}{B}}
\newcommand{\vz}{\stackrel{\circ}{V}}
\newcommand{\Vec}{\stackrel{\longrightarrow}}
\newcommand{\ddd}{\stackrel{{...}}}

\newcommand{\ov}{\overline}

\newcommand{\vp}{\varphi}

\newcommand{\bms}{\boldsymbol}

\newcommand{\Es}{E^star}

\vskip 1.5 true cm

\centerline{\bf About the Keplerization of motion}
\centerline{\bf in any central force field}

\vskip 0.5 true cm

\vskip 0.75 cm
\centerline{{\bf Christian CARIMALO}\footnote{{\bf christian.carimalo@sorbonne-universite.fr}}}

\vskip 0.75 cm

\centerline{Sorbonne Universit\'e, Campus Pierre et Marie Curie}
\centerline{4 Place Jussieu, 75005 Paris, France}

\vskip 1.5cm
\centerline{ABSTRACT}

\vv \nin The method of keplerization of one-body motion in any central force field, introduced by Martinusi and Gurfil in 2012, is reviewed and reformulated into a general ``homogenization" method which applies to any kind of bounded motion. It is also shown how this extended method provides a proof of the existence of a dynamical symmetry group and how it can be used to extend that group to a global symmetry group, for any such system.   

\vskip 0.75 cm

 
\vskip 0.75cm

\section{\large Introduction} 

\vv \nin In 2012, V. Martinusi and P. Gurfil introduced a clever method providing a link between any bounded (one-body) motion in a central force field and a similar keplerian one, justifying calling it a keplerization of such motion, Ref. \cite{Margur}. We will describe their method differently and extend it, rather focusing on geometrical trajectories, according to the point of view developed throughout this article.   

\vv \nin The system considered here is that of a classical and non-relativistic point particle $P$ of mass $m$, moving in an inertial frame having as origin the source $O$ of a spherical symmetric potential acting on $P$.  We will use the following notations : ${\bms r}$ is the position vector of $P$ relative to $O$, and $r = ||{\bms r}||$ ; ${\bms  v}$ and ${\bms p}= m{\bms v}$ are its velocity and momentum vectors, respectively, and $v = ||{\bms v}||$, $p=||{\bms p}|| = mv$.

\vv \nin Thanks to the spherical symmetry of the potential, the angular momentum of the particle, ${\bms L} = {\bms r} \times {\bms p}$, is a conserved vector, and consequently its motion corresponding to a given value of ${\bms L}$ takes place in a plane perpendicular to the latter and containing $O$. As the orientation of axes can be freely chosen, $Oz$ is usually taken along ${\bms L}$, the plane of motion being then the plane $(x, y)$. The position of $P$ in that plane may be defined as usual by the distance $r =OP$ and the polar angle $\theta$ of ${\bms r}$ relative to the axis $Ox$. Since $z=0$, $p_z=0$, only the component $L_z$ of the angular momentum is non-zero, and we have  

\beq L_z =  ||{\bms L}|| = L= m r^2 \dot{\theta} = {\rm const} \label{aires} \label{mtcin} \enq

\vv \nin from which the Kepler's second law (law of equal areas) is deduced. In the following, we exclude the case $L=0$ corresponding to motions along straight lines passing through $O$. Since the potential $V(r)$ does not depend explicitly on time, the Hamiltonian (or energy) of $P$, $ H = \di{ p^2 \over {2m}} + V(r)$, is also a conserved quantity $E$. The radial and orthoradial components of ${\bms p}$ being respectively $p_r = m \dot{r}$ and $p_\theta = m r \dot{\theta} = L/r$, we reexpress this Hamiltonien as 

\beq H = \di{{m {\dot{r}}^2}\over 2} + U(r)= E,~~{\rm where}~~U(r) = \di{L^2\over{2m r^2}} + V(r) \label{hamiltonien} \enq 

\vv \nin is the effective potential. Writing 
$dH/dt =0$, the distance $r$ is found to vary in time according to the equation 

\beq  \dot{p}_r = m \ddot{r} = - \di{{dU}\over{dr}} \label{eqr} \enq 

 \vv \nin From Eq. (\ref{hamiltonien}), permissible motions of $P$ satisfy 

$$ \di{p^2_r \over{2m}} = E-U(r) \geq 0 $$ 

\vv \nin Turning points are those for which $p_r =0$ or $U(r) =E$. Such points are inevitably  present in closed orbits but may also exist for non-closed and unbounded motions. At these points, $p_r$ changes sign. The modulus and the polar components of the velocity are  

$$ v = \sqrt{2m \left(E - V(r) \right)},~~v_\theta  = r\dot{\theta} = L/(m r),~$$
\beq  v_r =\dot{r} = \pm \sqrt{2m(E -V(r)) - L^2/(mr^2)}  \label{vitesse} \enq 

\vv \nin the sign of $v_r$ changing at turning points only ; at given $E$ and $L$, they depend on the sole variable $r$. From Eq. (\ref{mtcin}), $v_r$ is also linked to the derivative of the polar equation $r(\theta)$ since  

\beq \di{{dr}\over{d\theta}} = \dot{r}/\dot{\theta} = \di{{mr^2 \dot{r}}\over L} \label{polareq} \enq

\vv \nin Using Eqs. (\ref{vitesse}) and (\ref{polareq}), the polar angle of the radius vector ${\bms r}$ relatively to some initial position ${\bms {OP}}_0 = r_0 \,{\bms e}_{r0}$, can be expressed as

\beq \theta = \theta(r) = \theta_0 + \sqrt{\di{L^2 \over{2m}}} \, \di{\int^r_{r_0}} \di{{ \pm dR}\over{R^2 \sqrt{E-U(R)}}} \label{teta}  \enq   

\vv \nin where $\theta_0= \theta(r_0)$ and where the sign of the integrand should be chosen such that $\dot{\theta} = L/(m r^2)$ is always strictly positive : during its motion, the particle $P$ tends to rotate around $O$ in the same direction of rotation. In principle, inverting Eq. (\ref{teta}) gives $r$ as a function of $\theta-\theta_0$, $r_0$, $E$ and $L$, that is, what is usually called the polar equation. At this stage, that relation which does not involve the parameter time anymore, describes analytically the curve, usually called the trajectory, obtained by collecting all positions taken by the particle $P$ during its motion, with the prescribed initial condition. At given $E$ and $L$, this condition generally restricts the extent of the said curve. For example, it may happen that the trajectory does not encounter any turning point whereas this turning point is present on another trajectory with the same values of $E$ and $L$ but with a different initial condition. To clarify this point, we will consider the celebrated example of the Kepler problem. Before, for convenience, we will introduce a Binet-type formula.    
From Eqs. (\ref{polareq}), (\ref{vitesse}), and defining $u = 1/r$, we find 

$$ u^\prime = \di{{d u} \over{d \theta}}  = - \di{1 \over r^2} \di{{dr}\over{d \theta}} = - \di{m \over L} \,\dot{r},~~~{\rm and}$$
\beq  u^{\prime 2}  = F(u) = - u^2 + \di{{2m}\over L^2} \left[ E - V(1/u) \right]  \label{Fu}  \enq

\vv \nin It is important to note that, at given $E$ and $L$, the last formula of Eq. (\ref{Fu}) does not depend explicitly on any particular initial condition and thus applies as well to the longest possible trajectory that can be found, simply prolongating the shorter ones by considering all possible permissible values of $\theta$. Obviously, the polar equation of this longest trajectory,  whose parameters must depend only on the constants $E$ and $L$, describes all the main characteristics of the trajectories that can be encountered. In particular, it must account for the presence of turning points, inevitable as will be seen below. Viewed as a geometrical curve independently of any motion, this ``complete trajectory" is associated with an infinity of solutions, each corresponding to a particular initial position of the particle moving on this curve. 
Enlarging the point of view, we may consider that two motions are equivalent if they yield the same values of the constants $E$ and ${\bms L}$ that determine the parameters of the said complete trajectory (including the plane of motion), the latter being viewed as representative of the class of motions taking place on it.  From now on, we will focus on these complete geometrical curves and what is called a trajectory will be considered as one of them.

\section{\large Classical examples of trajectories} \label{extraj}

\subsection{ Newtonian potential $V(r) = -K/r,~K>0$}  

\vv \nin This is the attractive potential of the Kepler problem. From Eq. (\ref{Fu}), we have  

$$ F(u) = - u^2 + \di{{2m E}\over L^2} + \di{{2m K}\over L^2} u = (u_M - u)(u-u_m) ~~~{\rm with} $$ 
\beq u_{M,m} = \di{{mK}\over L^2} \left[ 1 \pm \sqrt{ 1 + \di{{2 E L^2}\over{m K^2}}} \right] \label{kep1} \enq 

\vv \nin The function $F(u)$ can be positive and its zeros $u_{M,m}$ are real only if $E > - \di{{m K^2}\over{2 L^2}}$, which is here the condition for the existence of motions. 

\vv \nin Differentiating the last formula of Eq. ({\ref{Fu}) with respect to $\theta$, we here obtain the equation : 

\beq u^{\prime \prime} = -u + \di{{mK}\over L^2} \enq 

\vv \nin whose solutions are given by 

$$ u = \di{{mK}\over L^2} + a \cos \theta + b \sin \theta $$

\vv \nin $a$ and $b$ being constants that, for a particular motion, are fixed by the initial conditions $u(\theta_0) = u_0$, $u^\prime(\theta_0) = u^\prime_0$, with $u^\prime_0 = \pm \sqrt{F(u_0)}$. Taking these conditions into account, we finally arrive at  

$$ u =  \di{{mK}\over L^2} +\left( u_0 -  \di{{mK}\over L^2} \right) \cos \psi + u^\prime_0 \sin \psi,~~~{\rm where}~~~ \psi = \theta- \theta_0 $$

\vv \nin the variations of $\psi$ being further subject to the condition $ u = 1/r >0$. The following cases are to be considered.

\subsubsection{ Case $E>0$  \label{kepEp}  }

\vv \nin Then, $u_m$ is negative, $u_M$ is positive and $F(u) = \left(u+|u_m|\right)\left( u_M -u \right) \geq 0$ only if $u \leq u_M$ or $r \geq r_m =1/u_M$. A turning point, for which $u^\prime = 0$ (see Eq. (\ref{Fu})), is possibly encountered during a motion, for $u=u_M$ and $\theta = \theta_t$. Since $\dot{\theta} >0$, this happens only if $\theta_t > \theta_0$. If $\theta_0 > \theta_t$, the particle goes to infinity without passing through the turning point. As the corresponding complete trajectory must contain this point, the latter can be chosen as a reference in order to fix the constants $u_0$ and $u^\prime_0$ (i.e., $\theta_0 = \theta_t$). Thus, taking $u_0 = u_M$, $u^\prime_0 =0$, and setting $\psi = \theta-\theta_t$, we obtain the well-known polar equation of a hyperbola of focus $O$   

$$ u = \di{{mK}\over L^2} \left[ 1+ e \cos \psi \right]~~{\rm where}~~ e =  \sqrt{ 1 + \di{{2 E L^2}\over{m K^2}}} > 1,~~{\rm or}$$
\beq r = \di{r_0 \over{ 1 + e \cos \psi}}~~{\rm with}~~r_0 = \di{L^2 \over{mK}} \label{hyperb} \enq

\vv \nin where $- \Psi < \psi < \Psi$ with $\Psi = \di{\pi \over 2} + \sin^{-1} \di{1 \over e}$. As expected, this equation does not make reference on any motion. 

\subsubsection{ Case $E=0$ \label{kepEz} } 

\vv \nin This is a limiting case of the previous one where $e=1$. The polar equation of the complete trajectory has the polar equation 

$$ r = \di{r_0 \over{1 + \cos \psi}}, ~~{\rm with}~~- \pi < \psi <+ \pi $$ 

\vv \nin which is that of a parabola of focus $O$. 

\subsubsection{ Case $- \di{{m K^2}\over{2 L^2}} < E<0$ \label{kepEm} } 

\vv \nin Here, $u_m$ and $u_M$ are both positive, and $F(u) = (u-u_m)(u_M -u)$ is positive if and only if $u_m \leq u \leq u_M$ : any motion takes place inside the crown defined by $1/u_M = r_m \leq r \leq r_M = 1/u_m$. We have turning points at $u=u_m$ and $u = u_M$. Proceeding as in the first case, we obtain for the polar equation of trajectories 

\beq r = \di{r_0 \over{ 1 + e \cos \psi}}~~{\rm with}~~e =\sqrt{ 1 - \di{{2 |E| L^2}\over{m K^2}}}   <1 \label{ellipse} \enq
$$ {\rm and}~~ r_m = \di{r_0 \over{1+e}},~~r_M = \di{r_0 \over{1-e}}  $$
 
\vv \nin which is that of an ellipse of focus $O$, with excentricity $e$. The motion is periodic, there is no limitation for $\psi$ and, in fact the particle runs indefinitely the said ellipse. We will then consider the ellipse as the unique representative trajectory for this case.

\subsection{\large Potential $V(r) = - K/r^2, ~K>0$}} 

\vv \nin For this attractive potential, we find 

\beq F(u) = - \alpha^2 u^2 + \di{{2mE}\over L^2} ~~{\rm and} ~~
 u^{\prime \prime} = -\alpha^2 u~~{\rm with}~~ \alpha^2 = 1- \di{{2m K}\over L^2},  \label{hom2} \enq

\vv \nin Assuming $L^2 \neq 2 m K$, we have the following cases. 
\vv

\subsubsection{ Case $\alpha^2 >0,~E >0$ \label{vm2Ep} }

\vv \nin Then, 

\beq F(u) = \alpha^2 (u^2_M - u^2)~~{\rm with}~~u_M = \sqrt{\di{{2mE}\over{\alpha^2 L^2}}} \label{hom2-1} \enq 

\vv \nin and we have a turning point at $u=u_M $. Motions are restricted to the domain $r \geq r_m = 1/u_M$ and the polar equation of the unbounded complete trajectory is found to be 

\beq  r(\psi) = \di{r_m \over \cos (\alpha \psi)}  \enq  

\vv \nin $\psi$ being the angle between ${\bms r}$ and its position at the turning point, varying in the interval $\left] - \di{\pi \over{ 2\alpha}}, + \di{\pi \over{ 2\alpha}} \right[$.

\subsubsection{ Case $\alpha^2 <0,~E < 0$ \label{vm2Em} }

\vv \nin Setting $\Omega^2 = - \alpha^2$, we have now 

\beq F(u) = \Omega^2 \left( u^2 - u^2_m \right)~~{\rm with}~~ u_m = \sqrt{ \di{{2m |E|}\over {\Omega^2L^2}}} \label{vm2Em-1} \enq 

\vv \nin and a turning point at $u = u_m$. Motions are bounded in the domain $r \leq r_M = 1/u_m$ and the polar equation of the complete trajectory is

\beq  r(\psi) = \di{r_M \over \cosh (\Omega \psi)}  \label{vm2Em-2} \enq  

\vv \nin $\psi$ being the angle between ${\bms r}$ and its position at the turning point,  here considered to vary in the interval $\left]-\infty, + \infty \right[$.

\subsubsection{ Case $\alpha^2 <0,~E >0$  } \label{notp}

\vv \nin This is the case where $r$ has no lower or upper limit, i.e. trajectories do not have any   turning point. When $u \rightarrow 0$, $u^{\prime} \rightarrow \epsilon \, \sqrt{\di{{2 m E}\over L^2}}$ with $\epsilon = \pm 1$, and the polar equation is found to be 

$$ u(\psi) = u_0 \sinh \left(\Omega \epsilon \psi \right) ~~{\rm or}~~ r(\psi) = \di{r_0 \over{\sinh \left(\Omega \epsilon \psi\right) }} ~~{\rm where} $$
\beq u_0 = 1/r_0 = \sqrt{\di{{2 m E}\over{\Omega^2 L^2}}},~\Omega = \sqrt{\di{{2 m K}\over L^2}-1} \label{eqnotp} \enq  

\vv \nin $\psi$ varying in the interval $\left]0, + \infty\right[$ or $\left]-\infty, 0 \right[$ according to whether $\epsilon = +1$ or $\epsilon =-1$. Note that here, the axis $Ox$ is not defined from a turning point but from the limit $r \rightarrow +\infty$. However, it is also a symmetry axis for the all set of trajectories of this domain. Note that Eq. (\ref{eqnotp}) can be obtained from Eq. (\ref{vm2Em-2}) by the substitution 

$$ r_M \rightarrow i \sqrt{\di{{2m E}\over{\Omega^2 L^2}}},~~~\Omega \psi \rightarrow \Omega |\psi| + i \di{\pi \over 2} $$

\subsection{\large Potential $V(r) = + K/r, ~K>0$  \label{antikep}} 

\vv \nin For this repulsive potential, $F(u) = -u^2 + \di{{2mE}\over L^2} - \di{{2m K}\over L^2} u$, and motions exist only for $E>0$. Trajectories are only lower bounded with a turning point at $u=u_M =( e-1)/r_0$ and $\psi=0$, where 

$$ r_0 = \di{L^2 \over{ mK}},~~ e= \sqrt{1 + \di{{2 E L^2}\over{m K^2}}} >1 $$

\vv \nin They are hyperbolas with polar equation 

\beq r(\psi) = \di{ r_m \over{-1 + e \cos \psi}}~~{\rm with}~~r_m = 1/u_M \enq

\vv \nin $\psi$ varying in the interval $\left] - \cos^{-1}(1/e), + \cos^{-1}(1/e) \right[ $.

\subsection{\large Potential $V(r) = + K/r^2, ~K>0$} 

\vv \nin This is another repulsive potential for which $F(u) = - u^2 \Omega^2 + \di{{2m E}\over L^2}$ with $\Omega^2 = 1 + \di{{2 m K}\over L^2} >1$. Motions exist only for $E>0$ and any trajectory is lower bounded at a turning point located at $u= u_M = \sqrt{\di{{2m E}\over{\Omega^2 L^2}}}$ and $\psi=0$, $(F(u) = \Omega^2 \left(u^2_M - u^2 \right)$).  The polar equation is 

\beq r(\psi) = \di{r_m \over{ \cos \Omega \psi}} ,~~{\rm with}~~ r_m = 1/u_M \enq

\vv \nin $\psi$ varying in the interval $\left] - \di{\pi \over{2 \Omega}}, + \di{\pi \over{2 \Omega}} \right[$.

\section{\large The existence of turning points and consequences} 

\vv \nin Looking at the entire variations of the function ${\cal E}(u) = \di{L^2 \over{2m}} u^2 + V(1/u)$ for a given $L$, it is obvious that we can always find a value of $E$ satisfying the equation $E = {\cal E}(u)$, equivalent to $F(u) =0$, in some domain of variations of $u$. Hence, whatever the shape of the potential, turning points do exist. Examples have been shown in the previous section. We have the following basic situations according to the kind of variation of $F(u)$ in the vicinity of the value $u_t$ of $u$ at a turning point. 

\vv \nin (i) According to whether $F(u)$ is increasing or decreasing from $u=u_t$, motions are possible only for $u \geq u_t$ or $u \leq u_t$. In the first case, $u_t$ is a minimum value $u_m$ whereas in the second case it is a maximum value $u_M$. 

\vv \nin (ii) For a given value $E$, $F(u)$ has two zeros $u_1$ and $u_2$ with $u_1 \leq u_2$ and no other zero in the interval $\left[u_1, u_2 \right]$. If $F(u) \leq 0$ for $u_1 \leq u \leq u_2$, motions are possible only in the domains defined by $u \leq u_1$ or $u \geq u_2$. Respectively to these domains, $u_1$ is a maximum value and $u_2$ is a minimum value, as described in (i). If $F(u) \geq 0$ in $\left[u_1, u_2 \right]$, $u_1$ and $u_2$ are a minimum $u_m$ and a maximum $u_M$. 

\vv \nin Consequently and accordingly to these cases, we can write the corresponding $F(u)$ in the form 

$$ F(u) = g_1(u) \left( u -u_m \right),~~F(u) = g_2(u) \left(u_M -u \right), $$
\beq F(u) = g_3(u) \left(u_M -u \right) \left(u- u_m \right) \label{lesfus} \enq

\vv \nin where the functions $g_i(u)$ ($i=1,2,3$) are strictly positive in the domain of motions and non-zero at the corresponding turning point(s). In the following, $\psi$ is the polar angle of ${\bms r}$ relative to its position at a turning point chosen to define the axis $Ox$.

\subsection{Keplerization for twice-bounded motions \label{kepdb}} 

\vv \nin Let us focus first on the third form written in Eq. (\ref{lesfus}) for motions that are twice-bounded, i.e. with a lower bound $r_m = 1/u_M$ and an upper bound $r_M =1/u_m$. From (\ref{Fu}) we have  

\beq u^{\prime 2} = g_3(u) \left(u_M -u \right) \left(u- u_m \right) \label{db1} \enq 

\vv \nin which expression is very similar to that obtained in the third case of the Kepler problem (subsection \ref{kepEm}). Then, noticing that $g_3(u)$ is dimensionless, it is very tempting to introduce a new angular variable $\chi$ such that 

\beq \di{{d \chi}\over{d \psi}} = \sqrt{g_3(u(\psi)) } \label{kzdb1} \enq

\vv \nin in order to transform Eq. (\ref{db1}) into the equation  

\beq \left(\di{{du}\over{d \chi}} \right)^2 =\left(u_M -u \right) \left(u- u_m \right) \label{kzdb2} \enq

\vv \nin which is identical to that of the Kepler problem in the case $u_m \leq u \leq u_M$. Thus, remarkably, the simple substitution of $\psi$ by $\chi$ allows us to make a link between a doubly bounded trajectory possibly provided by a potential $V(r)$ and an ellipse of the Kepler problem. This is the {\it keplerization} proposed by V. Martinusi and P. Gurfil. Let us note $E^\star$, $L^\star$ and $K^\star$ the energy, the angular momentum and the potential constant of the said Kepler problem ($V^\star(1/u) = - K^\star u$), respectively. From (\ref{kep1}) we must have  

\beq u_M + u_m = \di{{2 m K^\star} \over{L^{\star 2}} },~~u_m u_M = - \di{{2 m E^\star}\over{L^{\star 2}}} \label{cstkep} \enq 

\vv \nin while $u_m$ and $u_M$ verify  

\beq u^2_m = \di{{2m}\over L^2} \left[ E - V(1/u_m) \right],~~u^2_M = \di{{2m}\over L^2} \left[ E - V(1/u_M) \right] \label{ulim1} \enq 

\vv \nin In this keplerization, we must also allow a change of time : $t \rightarrow t^\star$. Then, 

$$ L^\star = m r^2 \di{{d \chi}\over{d t^\star}} = m r^2 \di{{d \chi}\over{d \psi}} \di{{d \psi}\over{dt}} \di{{dt}\over{d t^\star}} = L \sqrt{g_3(u)} \di{{dt}\over{dt^\star}}, $$

\vv \nin and since $L$ and $L^\star$ must be constant, $t$ and $t^\star$ must be correlated by 

$$ \di{{d t^\star}\over{dt}} = {\rm const.} \sqrt{g_3(u)} ~~~({\rm const.} = L/L^\star) $$

\vv \nin Note that $u_m$ and $u_M$ are here fixed by the values of $E$ and $L$, and the constraint $|E^\star| <  \di{{m K^{\star 2}}\over{2 L^{\star 2}}} $ (see subsection \ref{kepEm}), is already satisfied, accounting that $(u_m +u_M)^2 > 4 u_m u_M$. Then, from Eqs. (\ref{ellipse}) and (\ref{cstkep}), we have 

\beq r = \di{r^\star_0 \over{ 1 + e^\star \cos \chi}}~~{\rm with}~~e^\star =\di{{u_M - u_m}\over{u_M + u_m}} ,~~ r^\star_0 = \di{2 \over{u_M + u_m} } \enq

\vv \nin Only in the case where $g_3(u)$ is a (positive) constant, we can simply take $t^\star = t$, as shown by the following example. Consider the inhomogeneous potential 

\beq V(r) = -\di{K_1 \over r} + \di{K_2\over r^2},~~K_1, K_2 >0 \label{inhom}  \enq

\vv \nin We have 

$$ F(u) = -u^2 \left[ 1 + \di{{2m K_2}\over L^2} \right]+ \di{{2m E}\over L^2 } + \di{{2m K_1}\over L^2} u = \beta^2 \left(u_M -u \right)\left(u-u_m \right) ~~{\rm with} $$ 
\beq \beta^2 = 1 + \di{{2m K_2}\over L^2} ,~~ u_{M,m} = \di{{mK_1}\over {L^2 \beta^2}} \left[1 \pm \sqrt{1 + \di{{2 E L^2 \beta^2}\over{m K^2_1}}} \right] \enq

\vv \nin assuming here $- \di{{mK^2_1}\over{2 \beta^2 L^2}} < E<0$. Hence, $g_3(u) = \beta^2 = {\rm const}$. Thus, we can simply take $\chi = \beta \psi$ (then, $t^\star =t$, $E^\star =E$  and $L^\star = \beta L$), from which it results that  

\beq r = \di{r^\star_0 \over{ 1 + e^\star \cos \beta \psi}}~~{\rm with}~~e^\star =\sqrt{1 + \di{{2 E L^2 \beta^2}\over{m K^2_1}}} <1 ,~~ r^\star_0 = \di{{L^2 \beta^2}\over{mK_1}} \label{polV2} \enq

\vv \nin This example is instructive for two reasons. First, it shows that the keplerization can link an inhogeneous potential to an homogeneous one. Second, if $\beta$ is not a rational number, the twice-bounded trajectories provided by the potential Eq. (\ref{inhom}) are not closed and  the various positions taken by the particle during its motion fill the entire circular crown $r_m \leq r \leq r_M$. The keplerization provides a way to follow this complicated motion on a single ellipse.   

\subsection{Keplerization for lower-bounded motions \label{keplb}} 

\vv \nin In the case of lower-bounded motions ($r \geq r_m$), we have $F(u) = g_2(u) \left(u_M -u \right)$ ($u_M = 1/r_m)$. Obviously, $g_2(u)$ has the same dimension as $u$. In view of a keplerization, we will thus rewrite $g_2$ in the form 

\beq g_2(u) = G_2(u) \left(u + a \right) \enq 

\vv \nin where $G_2(u)$ is dimensionless and the constant $a$ is assumed to be stricly positive to avoid an additional singularity. To be closer to the form of $F(u)$ obtained in the Kepler problem with positive energy (subsection \ref{kepEp}), we will take $a < u_M$ and define $|u_m| =a$ with $u_m <0$. Then, introducing a new angle $\chi$ and a new time $t^\star$ such that 

$$ \di{{d \chi}\over{d \psi}} = \sqrt{G_2(u)} ,~~\di{{d t^\star}\over{dt}} = \di{L \over L^\star} \sqrt{G_2(u)} $$  

\vv \nin we obtain the equation 

\beq \left(\di{{du}\over{d \chi}} \right)^2 =\left(u_M -u \right) \left(u- u_m \right) \enq

\vv \nin leading to the polar equation $r(\chi)$ of an hyperbola, see Eq. (\ref{hyperb}),

\beq r = \di{r^\star_0 \over{ 1 + e^\star \cos \chi}}~~{\rm with}~~e^\star = \di{{u_M+|u_m|}\over{u_M - |u_m|}},~~r^\star_0 = \di{2\over{u_M-|u_m|}}  \label{kephyperb} \enq

\vv \nin where $- \Psi < \chi < \Psi$ with $\Psi = \di{\pi \over 2} + \sin^{-1} \di{1 \over e^\star}$. With this keplerization, any lower-bounded trajectory obtained from any potential $V(r)$ can be projected onto an hyperbola of the Kepler problem. 

\vv \nin Note however that this keplerization is not the only way to link this kind of trajectories to similar ones due to an homogeneous potential. First, we would have obtain a formula similar to Eq. (\ref{kephyperb}), using the Newtonian repulsive potential $V(r) = +K^\star/r$ ($K^\star>0$). Second, consider Eq. (\ref{hom2-1}) of subsection \ref{vm2Ep} corresponding to the attactive potential $V(r) = - K^\star/r^2$ with $K^\star >0 $, in the case $E^\star >0$. Taking 

$$ g_2(u) = G_2(u) \alpha^2 \left(u_M + u \right) ~{\rm with}~\alpha^2 = 1 -\di{{2m K^\star}\over L^{\star 2}}  > 0,$$ 
$$ E^\star = \di{{u^2_M L^{\star 2} \alpha^2}\over {2m}},~ \di{{d \chi}\over{d \psi}} = \sqrt{G_2(u)},  $$ 

\vv \nin yields to the polar equation 

\beq   r(\chi) = \di{r_m \over \cos (\alpha \chi)}~~{\rm with}~~r_m = 1/u_M \label{homm2} \enq 

\vv \nin In this last case, it would be more appropriate to speak of ``homogenization" rather than keplerization. Note also that a result similar to Eq. (\ref{homm2}) can be obtained using the repulsive potential $V(r) = + K^\star/r^2, K^\star >0$, with $\alpha = \sqrt{1 + 2m K^\star/L^{\star 2}}$, with the advantage that this does not imply any restriction on the value of $K^\star$. 

\vv \nin Let us finally remark that the above-mentionned case of subsection \ref{vm2Ep}, for which $F(u) = \alpha^2 (u^2_M - u^2)$, is even keplerizable, by setting  

\beq G_2(u) = \alpha^2 \di{{u_M + u}\over{|u_m| + u}},~\di{{d \chi}\over{d \psi}} = \sqrt{G_2(u)} \enq

\vv \nin This leads to the equivalent polar equations 

\beq r(\psi) =  \di{r_m \over \cos (\alpha \psi)} =   \di{r^\star_0 \over{ 1 + e^\star \cos \chi}}  \enq

\subsection{Homogenization for upper-bounded motions \label{UBM} } 

\vv \nin This is the case for which keplerization does not work because in the Kepler problem there is no motion having only an upper bound. In contrast, as shown in section \ref{vm2Em}, this situation exists with the attactive potential $V(r) = - K^\star/r^2, K^\star >0$ in the case $\Omega^2 = 2m K^\star/L^{\star 2} -1 >0$ and $E^\star < 0$. To match with this case, it is sufficient to redefine $g_1(u)$ as 

\beq  g_1(u) = \Omega^2 G_1(u) \left(u + u_m \right) ~~{\rm with}~~\Omega^2 >0, \enq  

\vv \nin where $G_1(u)$ is assumed to be stricly positive in the corresponding domain, to introduce a new angle $\chi$ such that 

\beq \di{{d \chi}\over{d \psi}} = \sqrt{ G_1(u)},\enq  

\vv \nin and finally to set

\beq K^\star = \di{L^{\star 2} \over{2m}} \left(\Omega^2 +1 \right), ~E^\star = - \Omega^2 u^2_m \di{L^{\star 2} \over{2m}} \enq 

\vv \nin Then, refering to Eqs (\ref{vm2Em-1}), (\ref{vm2Em-2}), we obtain the polar equation of trajectories in the form 

\beq r(\psi) = \di{r_M \over{\cosh \left(\Omega \chi(\psi) \right)}},~{\rm with}~ r_M = 1/u_m \enq      

\vv \nin $\chi$ varying in the interval $\left]-\infty, + \infty \right[$. Note, however, that the value of $K^\star$ depends on that of $L$.


\subsection{What about unlimited motions ? } \label{notp2} 

\vv \nin It may happen that for some values of $E$ and $L$ the corresponding trajectories of a given problem do not present any turning point. An example is given in subsection \ref{notp}. There is no limitation on $r$ which then can vary in the entire interval $]0, + \infty[$. In this case a  keplerization of the motion is not possible. However, we must have in mind that such unlimited motion is the result of the integration of the evolution equation of the studied system for particular values of $E$ and $L$, this same equation whose integration for other values of the latter parameters yields KH-able trajectories, since turning points are inevitably present somewhere for any system. The consequence is that unlimited motions inevitably bear a trace (visible or hidden) of this fact, possibly through analytic continuation. In the cited example, the axis $Ox$ defined by a turning point for bounded trajectories becomes an asymptote for unlimited motions, but is in both cases a global symmetry axis for the all sets of trajectories. Finally, we may even take this example as a reference for homogenizing unlimited motions of any other system, by setting 

$$ F(u) = G_0 (u) \Omega^{\star 2} (u^2 + u^{\star 2}_0),~{\rm with}~G_0(u) >0,  ~\di{{d \chi}\over{d \psi}} = \sqrt{G_0(u)} = \di{L^\star \over L} \di{{dt^\star}\over{dt}},  $$      

\beq u^\star_0 = 1/r^\star_0 = \sqrt{\di{{2 m E^\star}\over{\Omega^{\star 2} L^{\star 2}}}},~\Omega^\star = \sqrt{\di{{2 m K^\star}\over L^{\star 2}}-1},  \enq  

$$ u(\chi) = u^\star_0 \sinh \left(\Omega^\star \epsilon \chi \right) ~~{\rm or}~~ r(\chi) = \di{r^\star_0 \over{\sinh \left(\Omega^\star \epsilon \chi \right) }} $$

\vv \nin $\epsilon$ being defined as in subsection \ref{notp}, and  $L^\star < \sqrt{2 m K^\star}$, $E^\star >0$. 

\section{Analysis of findings} \label{findings}

\vv \nin The possibility of linking trajectories produced by any central potential to those  of the Kepler problem or those obtained from the potential $\propto ~-1/r^2$ is obviously due to the existence of turning points, giving to those trajectories a kind of similarity that may seem rude but is actually deeper, as shown in the following.    

\vv \nin In the study of the motions of the particle $P$ under the action of a central potential $V(r)$, the keplerization or homogenization method allows to associate it with a fictitious particle $P^\star$ of the same mass, moving in the same plane $(x,y)$ under the action of a known homogeneous central potential $V^\star(r)$, namely, either the Kepler's one, $\propto~ -1/r$, or the potential $\propto~ - 1/r^2$.       

\vv \nin Let ${\bms OP}^\star = {\bms r}^\star$ be the position vector of $P^\star$ at time $t^\star$ with  

\beq {\bms r}^\star = r\, {\bms e}^\star_r, ~{\bms e}^\star_r = \cos \chi\, {\bms e}_x + \sin \chi~{\bms e}_y  \label{postar1} \enq

\vv \nin The momentum of $P^\star$ is 

\beq {\bms p}^\star = m \di{{d {\bms r}^\star}\over{ d t^\star}} = m \di{{d \chi}\over{d t^\star}} \di{{d {\bms r}^\star}\over{ d \chi}} \,= m \di{{d \chi}\over{d t^\star}} \, \left[ \di{{dr}\over{d\chi}} {\bms e}^\star_r + r {\bms e}^\star_{\chi} \right] \label{mostar1} \enq
$$ ~{\rm with}~~ {\bms e}^\star_{\chi} = - \sin \chi\, {\bms e}_x + \cos \chi \,{\bms e}_y$$

\vv \nin and since $d \chi/d t^\star = L^\star/(mr^2)$, we have also ($u = 1/r)$

\beq {\bms p}^\star = L^\star  \left[ -\di{{du}\over{d \chi}}\, {\bms e}^\star_r + u \,{\bms e}^\star_\chi \right]  \enq 

\vv \nin The energy and angular momentum of $P^\star$ are, respectively,  

\beq E^\star = \di{{{\bms p}^{\star 2}}\over{2 m}} + V^\star(r),~{\bms L}^\star = {\bms r}^\star \times {\bms p}^\star = L^\star \, {\bms e}_z \label{ELstar} \enq 

\vv \nin They are obviously constant in time $t^\star$ as well as in time $t$, because, on one hand 

\beq \di{{d {\bms p}^\star}\over{ dt^\star}} =  - {\bms e}^\star_r \, \di{{d V^\star}\over{dr}},~~{\rm hence}~~\di{{d {\bms L}^\star} \over{d t^\star}} = {\bms 0} ,~~\di{{d E^\star} \over{d t^\star}} =0  \label{eqkep}  \enq 

\vv \nin and, on the other hand, 
$$\di{{d {\bms L}^\star}\over{dt^\star}} = \di{{d {\bms L}^\star}\over{dt}}  \di{{dt}\over{dt^\star}} ={\bms 0} ,~~\di{{d E^\star}\over{dt^\star}} = \di{{d E^\star}\over{dt}} \di{{dt}\over{dt^\star}}=0 ,$$
\beq {\rm hence}~~\di{{d {\bms L}^\star}\over{dt}} ={\bms 0},~~\di{{dE^\star}\over{dt}} =0 ,~~{\rm since}~~ \di{{dt}\over{dt^\star}}\neq 0 \label{constar1} \enq 

\vv \nin Beyond the fact that the keplerization-homogenization (KH) method can simplify the study of complicated motions due to complex potentials by following their evolutions along well known curves with polar equations $r(\chi)$ (see Ref. \cite{Margur}), it offers more with a better understanding of the dynamical symmetry that is inevitably attached to all these systems, especially about its very existence.  

\vv \nin Let us first consider the keplerization of section \ref{kepdb}. In this case, the particle $P^\star$ runs along ellipses under the action of the Kepler potential $-K^\star/r$ ($K^\star >0$). It is well known that this system has an additional first integral which is in fact a vector, namely the celebrated Laplace-Runge-Lenz (LRL) vector, Ref. \cite{LAP,RUN,LEN},

\beq {\bms A}^\star = \di{1\over m}\, {\bms p}^\star \times {\bms L}^\star - \di{K^\star \over r} {\bms r}^\star \label{larule} \enq

\vv \nin Since, again, $\di{{dt}\over{dt^\star}}\neq 0$, it is clear that the vector Eq. (\ref{larule}) is also a first integral for the motion of the particle $P$ : 

\beq \di{{d {\bms A}^\star}\over{d t^\star}} = {\bms 0} = \di{{d {\bms A}^\star}\over{d t}} \, \di{{dt}\over{d t^\star}}, ~~{\rm hence}~~ \di{{d {\bms A}^\star}\over{d t}} ={\bms 0}  \label{conslrl} \enq 

\vv \nin Let us specify what the LRL vector represents. A short calculation shows that 

\beq {\bms A}^\star = K^\star\, e^\star\, {\bms e}_x \label{larule2} \enq

\vv \nin where $e^\star$ is the eccentricity of the trajectory considered (ellipse, hyperbola or parabola). The unitary vector ${\bms e}_x$, which defines  the reference axis for both polar angles $\psi$ and $\chi$, is here chosen along the position vector ${\bms r}$ when the latter corresponds to a turning point, for which $dr/d\psi =0$ or $dr/d \chi=0$. In the Kepler problem when $E^\star <0$ (ellipses), there are only two turning points lying on a same straight 
line passing through $O$ and on either side of this point, one at $r=r_m$ (perihelion), the other at $r=r_M$ (aphelion). We take $\psi = 0, ~\chi=0$ for the perihelion, so that 
${\bms r}^\star = {\bms r}= r_m {\bms e}_x$ at this point. For the Kepler's hyperbolas or parabolas, only remains the perihelion and we will still choose ${\bms e}_x$ along the perihelion axis. Thus, in any case, ${\bms A}^\star$ is along the perihelion axis of Kepler's trajectories. Note that the emergence of a LRL vector for particle $P$ via its companion $P^\star$ has also been considered in Ref. \cite{Margur} but only for twice-bounded motions of $P$. 

\vv \nin From Bertrand's theorem, Ref. \cite{BERT}, we know that twice-bounded trajectories are all closed if and only if the potential is either the attractive Newtonian one $-K/r$ ($K >0$) or the attractive Hookean one $K^\prime r^2/2$ ($K^\prime >0$). For other potentials, the closure of such a trajectory can only be exceptional. In the general case, the motion fills the entire circular crown $r_m \leq r \leq r_M$ (see the example of section \ref{kepdb}) and we have an infinity of perihelion points with different angles. Since they are all equivalent, it is sufficient to choose one of them as a reference, which will also define the perihelion point of the associated Kepler's ellipse.

\vv \nin The conservation of the LRL vector is entirely due to the equation of motion Eq. (\ref{eqkep}) with $V^\star(r) = - K^\star/r$, independently of the nature of the trajectory, ellipse, hyperbola or parabola. Hence. from section \ref{keplb}, we infer that Eq. (\ref{conslrl})  also applies to any lower-bounded motion in its corresponding allowed domain. 
In particular, keplerizing the system with the attractive potential $V(r) = -K/r^2$ ($K>0$) in the case $E>0$, $2 m K < L^2$, see section \ref{vm2Ep}, we will find for it an associated conserved LRL vector Eq. (\ref{larule}), with appropriate definitions of ${\bms r}^\star$ and ${\bms p}^\star$. 

 \vv \nin Hence, for the above-mentionned cases, ${\bms e}_x$ is common to the system studied (with particle $P$) and its keplerized version (with particle $P^\star$) and is a first integral for both. Using either 

\beq  {\bms e}_x =  \cos \psi\, {\bms e}_r - \sin \psi\, {\bms e}_\psi,~~{\bms e}_r = {\bms r}/r,~~{\bms e}_\psi =  {\bms e}_z \times {\bms e}_r =  {\bms L}\times {\bms r}/(Lr)   \label{vr1} \enq

\vv \nin or analogous formulas for the keplerized version, we can express ${\bms e}_x$ in two equivalent forms  

\beq {\bms e}_x = \di{1 \over{L r}} \left[ \sin \psi \,{\bms r}\times {\bms L} + L \cos \psi \, {\bms r} \right] = \di{1 \over{L^\star  r}} \left[ \sin \chi \,{\bms r}^\star\times {\bms L}^\star + L^\star  \cos \chi \, {\bms r}^\star \right] \label{2forms}  \enq

\vv \nin $\psi$ and $\chi$ being now considered as functions of the rotational invariants $r$, $E$ and $L$ through the corresponding polar equations. In this way, ${\bms e}_x$ may be considered as a field vector in phase space. 

\vv \nin Let us remark that when expressed in terms of canonical variables, first integrals that do not depend explicitly on time owe their status to the sole fact that their Poisson brackets with the Hamiltonian are zero : viewed in this way, they are not attached to the class of trajectories from which they have been highlighted, they apply to all possible motions. In particular, having found the  conserved vector ${\bms e}_x$ for the potential $-K/r^2$ in the case of lower-bounded trajectories, that vector is also a first integral for the upper-bounded trajectories given by that potential. As the latter serves to homogenize the upper-bounded motion due to  any other potential, as decribed in section \ref{UBM}, we infer that this first integral is also present for any of these motions. A similar remark can be made for trajectories without turning point (see  subsections \ref{notp} and \ref{notp2}) and where keplerization does not apply.  

\vv \nin From all these considerations, we conclude that, in additon to the classical first integrals energy and angular momentum, any one-body motion in a central force field has another  independant first integral which is a vector that do not depend explicitly on time.  

\vv \nin By the way, note that, contrary to some belief, the existence of said additional first integral is in no way related to that of closed trajectories. A clear evidence of this fact is given by the example of the repulsive Newtonian potential $V(r) = + K/r$ with $K>0$, which provides only hyperbolas and parabolas (lower-bounded trajectories) and a LRL-like vector 

\beq {\bms A} = \di{1\over m}\, {\bms p} \times {\bms L} + \di{K \over r} {\bms r} \label{LRL2} \enq

\vv \nin which is a first integral. Here again, this vector can be expressed as in Eq. (\ref{larule2}), 
with ${\bms e}_x$ along the perihelion axis. 

\vv \nin The existence of additional first integrals is commonly thought at the origin of a dynamical symmetry. What is its nature in the present case ? Obviously, it has something to do with the vector ${\bms e}_x$. Since that  vector is defined by turning points, the latter appear essential in the existence of LRL-like vectors. This is detailed in Ref. \cite{cari3}, where it is shown how their importance comes basically from the spherical symmetry of the studied systems, which (i) makes the distance $r$ the only remaining parameter determining the properties of said systems, primarily the existence and location of turning points, (ii) makes these points define axes of symmetry of trajectories. 

\vv \nin This symmetry of trajectories is actually the minimal common property of all the systems under study. It is at the origin of a common dynamical symmetry, which is expressed in the existence of a continuous group of transformations, the dynamical symmetry group. 

\section{The dynamical symmetry} 

\subsection{The dynamical symmetry group}

\vv \nin Based on the examples of the dynamical symmetry groups found in the Kepler and the Hooke problems, various authors have already searched for analogous symmetry groups that could be associated with other central force fields, see e.g. Refs \cite{FRD,MUK,BRS}. However, to our knowledge, neither the very reason for the inevitable existence of such a group nor its actual action on trajectories have been as well clarified as what we have tried to do in Ref. \cite{cari3}. The KH method here presented provides further evidence of the existence of this group in a very beautiful, perhaps best way, with a breakthrough in its physical significance and a perspective of further developments. As shown in Ref. \cite{cari3}, it appears that the structure of said group can be described by a universal formalism, whatever the potential. Let us define the LRL-like vector 

\beq {\bms S} = L\, {\bms e}_x \label{symvec} \enq

\vv \nin called the {\it symmetry vector}, ${\bms e}_x$ being, in the current case, the unitary vector of a perihelion axis. The symmetry vector is of course a first integral and verifies the set of relations 
involving Poisson brackets 

\beq \{L_i, S_j \}= \ep_{ilk} S_k,~~\{S_i, S_j\}= - \ep_{ijk} L_k,~~\{L_i, L_j\} = \ep_{ijk} L_k \label{lieal} \enq

\vv \nin defined as follows 

\beq \{ \Phi_1, \Phi_2\} = \di{\sum^3_{k=1}} \left[ \di{{\partial \Phi_1}\over{\partial x_k}} \di{{\partial \Phi_2}\over{\partial p_k}} -\di{{\partial \Phi_1}\over{\partial p_k}} \di{{\partial \Phi_2}\over{\partial x_k}}  \right] \label{poisson1} \enq

\vv \nin $\Phi_1$and $\Phi_2$ being two arbitrary functions of the canonical variables $x_k$ and $p_k$ ; in Eq. (\ref{lieal}), $\epsilon_{ijk}$ is the 3-rank Levi-Civit\`a tensor with $\epsilon_{123}=1$ and summation on repeated indices is assumed. As operators by means of Poisson brackets, the six components $L_i$ and $A_i$ generates the dynamical symmetry group which, according to the structure of Eq. ({\ref{lieal}), is homomorphic to the groups $SO(3,1)$, or the Lorentz group $L(3,1)$ or $SL(2,C)$, see e.g. Ref. \cite{BAC}.

 \subsection{The complete symmetry group} 

\vv \nin From Eq. (\ref{Fu}) and the numerous examples given in section \ref{extraj}, it is clear that the parameters of trajectories of a same species and lying in a same plane are completely determined by the values of $E$ and $L$ (the inherent constants of potential being set aside). Thus, moving from one trajectory to another of the same species is equivalent to changing the values of $E$ and $L$ in the permitted range corresponding to this species. Changing the plane of trajectories and the value of $L$ can be done by the dynamical symmetry group discussed above. However, since its generators are first integrals, they have zero Poisson brackets with the Hamiltonian and consequently, said group fails to change the value of $E$.    

\vv \nin This failure has already been pointed out in Ref. \cite{cari1} when studying the cases of the attractive Newtonian and Hookean potentials. In that reference, noticing that these two potentials are homogeneous, we have proposed to use the concept of {\it mechanical similarity} to cure this problem. 

\vv \nin Consider a potential $V(r)$ that is an homogeneous function of degree $\nu$ : if $r \rightarrow a\,r$, then $V(r) \rightarrow a^\nu V(r)$. The equations of motion are invariant under the substitution 

\beq x_k \rightarrow a \, x_k,~~t \rightarrow t\, a^{1 - \nu/2}~~{\rm or}~~p_k \rightarrow p_k a^{\nu/2}  \enq

\vv \nin In phase space, the infinitesimal generator of this transformation is 

\beq {\cal M} = x_k \di{\partial \over{\partial x_k}} + \di{\nu \over 2}\, p_k \di{\partial \over{\partial p_k}} \label{opM} \enq

\vv \nin  where summation on repeated indices is assumed. The action of this operator on the Hamiltonian $H = \di{{\bms p}^2\over{2m}} + V(r)$, on the componenta of the angular momentum $L_i = \epsilon_{ijk} x_j p_k$ and on the magnitude $L$ of the latter are 

\beq {\cal M}(H) = \nu H,~~{\cal M}(L_i) = \left( 1 + \di{\nu \over 2} \right) L_i,~~{\cal M}(L) = \left( 1 + \di{\nu \over 2} \right) L  \label{actM} \enq   

\vv \nin Defining the operator ${\cal N}(Q) = \{ N, Q\}$ for any functions $N$ and $Q$ of canonical variables, we have established the relations with commutators, see Ref. \cite{cari1} : 

$$ \left[ {\cal M}, {\cal L}_i \right] = \left[ {\cal M}, {\cal S}_i \right] =0,~~~ \left[ {\cal L}_i, {\cal L}_j \right] = \epsilon_{ijk}\, {\cal L}_k $$
\beq \left[ {\cal L}_i, {\cal S}_j \right] = \epsilon_{ijk}\, {\cal S}_k,~~ \left[ {\cal S}_i, {\cal S}_j \right] = -\epsilon_{ijk}\, {\cal L}_k \label{alsym} \enq

\vv \nin where ${\cal S}_i = \{ S_i, \,\cdot \, \}$ and 

$$ {\cal L}_i (Q) = \{L_i, Q\} = {\cal L}_{x i}(Q) + {\cal L}_{p i}(Q),~~~{\rm with} $$ 
$$ {\cal L}_{x i} = - \epsilon_{ijk} x_j \di{\partial \over{\partial x_k}},~~~{\cal L}_{p i} = - \epsilon_{ijk} p_j \di{\partial \over{\partial p_k}} $$

\vv \nin The relations (\ref{alsym}) are that of the Lie algebra generating the {\it complete symmetry group} of the problem with the homogeneous potential $V(r) \,\propto\, r^\nu$. Not only that group links trajectories of the same species lying in different planes with different values of $L$, it also links trajectories of the same species with different values of $E$. Note also that these relations are independent of the degree $\nu$, and that ${\cal M}(L_i) =0$ for $\nu =-2$. 

\vv \nin Amazingly, the KH method not only provides a proof of the existence of a dynamical symmetry group, it also offers the possibility to enlarge this group to a wider symmetry group, since it  connects trajectories due to any (spherical symmetric) potential to those obtained with homogeneous potentials, for which mechanical similarity applies. Obviously, if the potential is itself homogeneous, the Lie algebra of the associated complete symmetry group is already defined by Eqs. (\ref{alsym}) and (\ref{opM}). More spectacular is the case of inhomogeneous potentials. Assuming both angles $\psi$ and $\chi$ to vary continuously in the greatest possible interval (possibly $] - \infty, + \infty [$ even for closed trajectories), the relations 

\beq r^\star = r,~~\di{{d \chi}\over{d \psi}} > 0 \enq 

\vv \nin ensure a one-to-one correspondance between $P$ and its ``homogenized" companion $P^\star$ considered in Sec. \ref{findings}. The Lie algebra of the complete symmetry group for $P^\star$ is here again defined by equations (\ref{opM}) and (\ref{alsym}), where letters are just replaced with their ``starred" analogues. However, in this case, we face with the difficulty of linking  the new Poisson brackets involving the canonical variables of $P^\star$ with those involving the canonical variables of $P$ (defined in Eq. (\ref{poisson1})). This jeopardizes the possibility to express the action of the ``starred" symmetry group directly in terms of the variables associated with $P$. At this stage, it is only possible to described the action of said group on $E^\star$ and $L^\star$ and then derive its effect on $E$ and $L$ by inverting the transformation $P \rightarrow P^\star$. In this regard, it seems very difficult to determine whether the latter transformation is canonical or not, even in the framework of extended Hamiltonian Mechanics, see e.g. Ref. \cite{ODJ}. Actually, it is shown in Appendices A and B, that in a simple case, said transformation is not canonical. Despite this frustrating situation, the KH method allows us to prove the existence of a global symmetry group for the systems studied here, which was in fact the main goal of this work.

\section{Conclusion}

\vv \nin To deal with the problem of the one-body motion in a central force field, the KH method is an attempt to classify motions into groups in which they present some similarity, according to the response of their respective potentials for given values of energy and angular momentum. The effective existence of turning points whatever the shape of this force field is crucial for this  classification. It is known that turning points define axes of symmetry of trajectories and this feature is thus a common property of all motions. As shown in Ref. \cite{cari3}, this property originates a dynamical symmetry whose structure is universal, whatever the shape of the potential and the kind of motions, even motions that occur in regions where turning points are absent for some values of the parameters $E$ and $L$. The first merit of the KH method is to provide another proof of this fact, by linking said groups of motions to known solvable problems such as the Kepler problem with its celebrated LRL vector, or that with the potential $-K/r^2$ which itself can be linked in some way to the Kepler problem (see subsection \ref{keplb}). From a practical point of view, the re-parametrization of KH method allows  to follow complicated motions along simpler trajectories. The second merit of the KH method is to associate any motion with a motion in an homogeneous potential, for which the concept of mechanical similarity applies, namely either a motion of the Kepler problem or a motion in the potential $-K/r^2$.     

\vv \nin From our point of view, finding a continuous dynamical group for the systems considered in this article appears rather natural. An obvious but essential condition for the existence of such a group is the possibility of a continuous (or analytic) link between trajectories. This possibility does exist for our systems, and said link is supposed to be realized mathematically by the transformations of the group. Changing a trajectory into another of the same species remains to change continuously the values of the fundamental first integrals, energy $E$ and angular momentum ${\bms L}$. But the latter being Poisson-involutive, the transformations they generate through Poisson brackets are unable to achieve this change. The rotational group also cannot change the values of $E$ and $L$. The only possible conclusion is that there must exist at least one additional first integral which has non-zero Poisson brackets with the fundamental involutive first integrals and thus generates transformations changing in particular the value of $L$. Moreover, icing on the cake, the KH method is well adapted to extend the dynamical group to a larger symmetry group linking all trajectories of a same species but with different values of $E$ and $L$, whose existence was in fact predictable. 

\vv \nin From this observation, it is also natural to ask whether a dynamical group could exist for any integrable system whose solutions can be linked continuously. If so, such an integrable system would  also be {\it de facto} superintegrable. Then, the simple condition of a continuous link between solutions would appear independent of and, conceptually, more important than the very nature of the system under study, that nature manifesting only in the representation of the group and the degree of superintegrabllity, specific to that system. This last conjecture is also motivated by the fact that the dynamical group considered in this article can be described by a single formalism, whatever the potential.

\newpage

\bibliographystyle{amsplain}

\renewcommand{\refname}{\large{References}}

 \newpage

\setcounter{equation}{0}
\renewcommand{\theequation}{\mbox{A.}\arabic{equation}}

\nin {\large \bf Appendix A : About the keplerization of the problem with $V(r) = - K_1/r + K_2/r^2$} \label{nocanon}

\vv \nin Let us first establish some formulas that will be useful to check whether the transformation $P \rightarrow P^\star$ is canonical when the potential is $V(r) = - K_1/r + K_2/r^2$. We will be helped by general formulas obtained in Appendix B to which we refer the reader.   

\vv \nin Reconsider the vectors defined in Eq. (\ref{vr1}), 

$$ {\bms e}_r = {\bms r}/r = \cos \psi ~{\bms e}_x + \sin \psi~{\bms e}_y,~~{\bms e}_\psi = {\bms \ell} \times {\bms e}_r  = - \sin \psi~{\bms e}_x + \cos \psi~{\bms e}_y,~~{\bms \ell} = {\bms L}/L $$

\vv \nin Since $e_{ri} = x_i/r$, we have obviously $\{ r, e_{ri}\} = 0$, and (see Appendix B) 

$$ \{ r, e_{\psi i} \} = \epsilon_{ij k} \{r, \ell_j e_{rk} \} = \epsilon_{ijk} \ell_j \{r, e_{rk} \} = 0 $$ 

\vv \nin Then, with $e^\star_{ri} = x^\star_i/r $, $e^\star_{\chi i}= \epsilon_{ijk} \ell_j e^\star_{rk} $, we obtain 

$$ e^\star_{ri} = \cos \chi~e_{x i} + \sin \psi~e_{y i} = c \, e_{ri} + s \, e_{\psi i} $$
$$ e^\star_{\chi i} =  - s \, e_{ri} + c \, e_{\psi i}~~{\rm where}~~
 c = \cos (\chi - \psi),~s= \sin (\chi - \psi) , ~{\rm and}$$
$$ \{r, e^\star_{ri}\} = e_{ri} \{r,c\} + e_{\psi i} \{r,s\} $$

\vv \nin The Poisson brackets $\{r,c\}$ and $\{r, s\}$ are expressed as follows 

$$ \{r, c\} = -s \, \left(\di{{\partial r}\over{\partial x_k}}\right)_{\bms p} \left[ \di{\partial \over{\partial p_k}} (\chi - \psi) \right]_{\bms r} ,~ 
 \{r, s\} = c \, \left(\di{{\partial r}\over{\partial x_k}}\right)_{\bms p} \left[ \di{\partial \over{\partial p_k}} (\chi - \psi) \right]_{\bms r} $$

\vv \nin hence,  

\beq  \{r, e^\star_{ri} \} = e^\star_{\chi i}  \di{x_k \over r}  \left[ \di{\partial \over{\partial p_k}} (\chi - \psi) \right]_{\bms r} \enq

\vv \nin From the polar equations $r(\chi)$ and $r(\psi)$, the angles $\chi$ and $\psi$ are obtained as functions of $r$, $E$ and $L$. Thus, 

$$  \left[ \di{\partial \over{\partial p_k}} (\chi - \psi) \right]_{\bms r} =  \left[ \di{\partial \over{\partial E}} (\chi - \psi) \right]_{r, L} \di{{\partial E}\over{\partial p_k}} +  \left[ \di{\partial \over{\partial L}} (\chi - \psi) \right]_{r,E} \di{{\partial L}\over{\partial p_k}} ,~{\rm with} $$
$$ \di{{\partial E}\over{\partial p_k}} = \di{p_k \over m},~\di{{\partial L}\over{\partial p_k}} = \di{1\over L} \left[ r^2 p_k - ({\bms p}\cdot {\bms r}) x_k \right] = r e_{\psi k}$$

\vv \nin and consequently 

\beq \{r, e^\star_{ri} \} = e^\star_{\chi i} \, \di{{{\bms p}\cdot {\bms r}}\over{ m\, r}}  \left[ \di{\partial \over{\partial E}} (\chi - \psi) \right]_{r, L} \label{poirers} \enq

\vv \nin Let us now concentrate on the Poisson brackets $\{x^\star_i, x^\star_j\}$. We have

$$ \{x^\star_i, x^\star_j\} = \{r e^\star_{ri}, r e^\star_{r j}\} = r \{ e^\star_{ri}, r e^\star_{rj} \} + e^\star_{ri} \{r, r e^\star_{rj}\} = r e^\star_{ri} \{r, e^\star_{rj} \} - r e^\star_{rj} \{ r, e^\star_{ri}\} $$

\vv \nin From Eqs. (\ref{poirers}) and (\ref{ort3}) we thus obtain 

\beq \{x^\star_i, x^\star_j\} = \epsilon_{ij k} \ell_k \, \di{{{\bms p}\cdot {\bms r}}\over{ m}}  \left[ \di{\partial \over{\partial E}} (\chi - \psi) \right]_{r, L}  \label{poisxsxs}  \enq

\vv \nin Generally, $\chi$ and $\psi$ do depend on $E$ and the second member of Eq. (\ref{poisxsxs}) does not vanish : the Poisson brackets $\{x^\star_i, x^\star_j \}$ are not zero. This sole fact is sufficient to state that the transformation $P \rightarrow P^\star$ is not canonical.  

\vv \nin Consider again the example given in Section \ref{kepdb} where the potential is $V(r) = - K_1/r + K_2/r^2$ with positive $K_1$ and $K_2$. In this case, we have 

$$\chi = \beta\, \psi,~t^\star=t, ~E^\star = E,~L^\star = \beta L,~{\rm with}~ \beta = \sqrt{ 1 + \di{{2m K_2}\over L^2}  } $$  

\vv \nin Since time is unchanged, this would be the case of a traditional canonical transformation. 
From Eq. (\ref{polV2}) we have 

$$ \cos \chi = (u r^\star_0 -1)/e^\star, ~ {\bms p}\cdot {\bms r} = m r \dot{r} = - r L^\star \di{{du}\over{d \chi}} = r L^\star \di{e^\star \over r^\star_0} \sin \chi ,$$ 
$$ {\bms p}\cdot {\bms r} \left[ \di{\partial \over{\partial E}} (\chi - \psi) \right]_{r, L} = \left(1 - \di{1\over \beta}\right)  r L^\star \di{e^\star \over r^\star_0} \sin \chi \,\left[ \di{{\partial \chi} \over{\partial E}}\right]_{r, L} = -  \left(1 - \di{1\over \beta}\right)  r L^\star \di{e^\star \over r^\star_0} \left[ \di{{\partial \cos \chi} \over{\partial E}}\right]_{r, L} $$
$$ =  \left(1 - \di{1\over \beta}\right) \di{{r L^\star} \over{ e^\star r^\star_0}} (u r^\star_0 -1) \left[ \di{{\partial e^\star} \over{\partial E}}\right]_L = (\beta-1) \di{L \over {K_1 e^\star}} r \cos \chi \neq 0$$

\vv \nin and even in this simple case, the keplerization is not a canonical transformation.

\newpage

\setcounter{equation}{0}
\renewcommand{\theequation}{\mbox{B.}\arabic{equation}}

\nin {\large \bf Appendix B : Poisson brackets between components of unitary vectors orthogonal to the angular momentum}

\vv \nin Let ${\bms u}$ be any unitary three-vector field depending on canonical coordinates ${\bms r}$ and ${\bms p}$, and orthogonal to the angular momentum ${\bms L}= {\bms r}\times {\bms p}$. It is worth reminding that ${\bms u}$ owes its status of vector as regards the rotational group generated by the components of ${\bms L}$ by means of Poisson brackets, because   

$$ \{ L_i, u_j \} =  \left[ \di{{\partial L_i} \over{\partial x_k}} \di{{\partial u_j} \over{\partial p_k}} - \di{{\partial L_i} \over{\partial p_k}} \di{{\partial u_j} \over{\partial x_k}}  \right] = \epsilon_{ijk} u_k$$

\vv \nin with implicit summation on repeated indices. The Poisson brackets $U_{ij} = \{u_i, u_j \}$ define an antisymmetric two-rank tensor $U$ which is orthogonal to ${\bms L}$, since  

\beq L_i \{u_i, u_j\} = \{ L_i u_i, u_j\} - u_i \{L_i, u_j \} = - \epsilon_{ijk} u_i u_k = 0 \label{ort1} \enq

\vv \nin It is also orthogonal to ${\bms u}$ because 

\beq u_i U_{ij} = \di{1\over 2} \{u_i u_i, u_j\} = \di{1\over 2} \{1, u_j\} = 0 \label{ort2} \enq 

\vv \nin Then, let us define ${\bms \ell} = {\bms L}/L$ with $L = ||{\bms L} ||$, and ${\bms w} = {\bms \ell} \times {\bms u}$. The three unitary vectors ${\bms u}, {\bms w}, {\bms \ell}$ form an  orthonormal basis with direct orientation and as such verify the relations  

$$ u_i u_j + w_i w_j + \ell_i \ell_j = \delta_{ij}  $$  
\beq u_i w_j - u_j w_i = \epsilon_{ij k} \ell_k,~~w_i \ell_j - w_j \ell_i = \epsilon_{ijk} u_k,~~\ell_i u_j - \ell_j u_i  = \epsilon_{ijk} w_k \label{ort3} \enq
$$ u_i = \epsilon_{ijl} w_j \ell_k,~~w_i = \epsilon_{ijk} \ell_j u_k,~~\ell_i = \epsilon_{ijk} u_j w_k $$

\vv \nin Writing 

$$ U_{ij} = \delta_{ia} U_{a b} \delta_{jb} = \left( u_i u_a + w_i w_a + \ell_i \ell_a  \right) U_{ab} \left( u_b u_j + w_b w_j + \ell_b \ell_j  \right) $$ 
$$= w_i \left(w_a U_{ab} w_b\right) w_j $$

\vv \nin where Eqs (\ref{ort1}) and (\ref{ort2}) have been used on both sides,  we obtain for $U_{ij}$ a symmetric expression while it has to be antisymmetric. We thus conclude that is is in fact zero. This conclusion holds also for the Poisson brackets of the components $w_i$. Thus, 

\beq \{u_i, u_j\} =0,~~\{w_i, w_j\} =0  \label{nul1} \enq

\vv \nin Let us now consider the tensor $T_{ij} = \{u_i, w_j\}$. On its left side it is orthogonal to ${\bms u}$, see Eq. (\ref{ort1}) ; and also to ${\bms w}$ because, from Eq. (\ref{nul1}),    

$$ w_i \{u_i, w_j \} = \{ w_i u_i, w_j \} - u_i \{w_i, w_j \} = - u_i \{w_i, w_j\} = 0 $$

\vv \nin On its right side, $T_{ij}$ is orthogonal to ${\bms w}$ and also to ${\bms u}$ : 

$$ \{ u_i, w_j\} u_j = \{u_i, w_j u_j\} - \{u_i, u_j \} w_j = 0 $$ 

\vv \nin Then, 

$$ T_{ij} = \delta_{ia} T_{a b} \delta_{jb} = \left( u_i u_a + w_i w_a + \ell_i \ell_a  \right) T_{ab} \left( u_b u_j + w_b w_j + \ell_b \ell_j  \right) $$
$$= \ell_i \left(\ell_a T_{ab} \ell_b\right) \ell_j $$

\vv \nin But 

$$ L_a \{u_a, w_b\} = \{L_a u_a, w_b\} - u_a \{L_a, w_b \} = - u_a \epsilon_{abc} w_c = \ell_b,~~{\rm and} $$ 
$$ \ell_a \{u_a, w_b \} = \ell_b/L $$ 

\vv \nin Hence, 

\beq \{ u_i, w_j \} = \ell_i \ell_j/L = L_i L_j/L^3 \enq 

\vv \nin Let us insist on the fact that all these relations are completely independent of the actual form of the vector ${\bms u}$ : they are valid for any unitary vector orthogonal to ${\bms L}$.

\end{document}